\documentclass[twocolumn,showpacs,preprintnumbers,amsmath,amssymb]{revtex4}

\usepackage{graphicx}% Include figure files
\usepackage{dcolumn}% Align table columns on decimal point
\usepackage{bm}% bold math

\def\rbc{RB$_2$C$_2$}
\def\labc{LaB$_2$C$_2$}
\def\tbbc{TbB$_2$C$_2$}
\def\hbc{HoB$_2$C$_2$}
\def\dbc{DyB$_2$C$_2$}
\def\cb{CeB$_6$}
\def\pb{PrPb$_3$}
\def\tn{$T_{\rm N}$}
\def\tq{$T_{\rm Q}$}

\def\kpb{{\boldmath $k$}$_2=(0\,1\,1/2)$}

\def\kpd{{\boldmath $k$}$_4=(0\,0\,1/2)$}

\def\rab{1\,1\,0}
\def\rba{1\,${\bar 1}$\,0}
\def\rc{0\,0\,1}
\def\ea{${\langle}1\,0\,0{\rangle}$}

\def\eab{${\langle}1\,1\,0{\rangle}$}

\def\Hdir{$H{\parallel}$}

\begin{document}

%%%%%%%%%%%%%%%%%%%%%%%% Title and Authors %%%%%%%%%%%%%%%%%%%%%%%%%%%%

\title{Magnetic Phase Diagrams with Possible Field-induced Antiferroquadrupolar Order in {\tbbc}}

\author{Koji Kaneko}
\email{kanekok@neutrons.tokai.jaeri.go.jp}
\altaffiliation[Present address: ]{Advanced Science Research Center, Japan Atomic Energy Research Institute, Tokai, Naka, Ibaraki 319-1195, Japan}
%Lines break automatically or can be forced with \\
\author{Hideya Onodera}%
\author{Hiroki Yamauchi}
\author{Takuo Sakon}
\author{Mitsuhiro Motokawa}
\author{Yasuo Yamaguchi}
\affiliation{%
Institute for Materials Research, Tohoku University, Sendai 980-8577, Japan
}%

\date{\today}

%%%%%%%%%%%%%%%%%%%%%%%% Abstract %%%%%%%%%%%%%%%%%%%%%%%%%%%%%%%
\begin{abstract}
Magnetic phase diagrams of a tetragonal antiferromagnet {\tbbc} were clarified 
by temperature and field dependence of magnetization.
It is noticeable that the N{\'e}el temperature in {\tbbc} is anomalously enhanced with magnetic fields,  
in particular the enhancement reaches 13.5\,K for the {\eab} direction at 10 T.
The magnetization processes as well as the phase diagrams are well interpreted assuming that 
there appear field-induced antiferroquadrupolar ordered phases in {\tbbc}.
The phase diagrams of the AFQ compounds in {\rbc} are systematically understood 
in terms of the competition with AFQ and AFM interactions.
\end{abstract}

%%%%%%%%%%%%%%%%%%%%%%%  PACS  %%%%%%%%%%%%%%%%%%%%%%%%%%%%%%%%

\maketitle

%%%%%%%%%%%%%%%%%%%%%%%% Main text  %%%%%%%%%%%%%%%%%%%%%%%%%%%%%

In addition to spin and charge, an orbital degree of freedom in $3d$ and $f$ electron systems invites an upsurge of interests, 
because the coupling of the degree of freedom induces novel and rich variety of physical properties. 
The strong spin-orbit coupling is characteristic to $f$electron systems, 
hence {\boldmath $J$} is the basis of the quantum mechanical description.
The magnetic order parameter can be described by a linear combination of {\boldmath $J$}$_{Z\,,\rm {\pm}}$.
Depending on the interactions, a higher order term could be the primary order parameter. 
Quadrupolar ordering may occur without any magnetic contribution.
The competitive coexistence of dipolar and quadrupolar interactions and 
their response to pressure and magnetic field induce novel magnetic phenomena.
  
Recently Yamauchi {\it et al.} reported the antiferroquadrupolar (AFQ) order in the rare earth compound {\dbc}\,{\cite{dy1}}
with the tetragonal {\labc}-type structure\,{\cite{RBCcrys_1,RBCcrys_2}}.
{\dbc} undergoes an AFQ transition at {\tq}=24.7\,K.
The AFQ order in {\dbc} was directly confirmed by resonant X-ray scattering technique\,{\cite{DyATS_1,DyATS_2}}.
Note that {\tq} in {\dbc} is about ten times higher than those of other AFQ materials found to date, 
though the origin of the strong AFQ interaction is still an open question. 
Below {\tn}=15.3\,K, the antiferromagnetic (AFM) ordering coexists with the AFQ order.
The similar coexistent phase has been reported in the isostructural compound {\hbc}\,{\cite{Ho_1,Ho_2}}. 
The AFQ order in the {\rbc} compounds is the first example in which AFQ order is realized in the tetragonal symmetry.

This unusually strong quadrupolar interaction can also be expected in another isostructural compound {\tbbc}. 
{\tbbc} is an antiferromagnet with {\tn}=21.7\,K\,{\cite{kk3}}. 
{\tbbc} shows an anomalous increase of the magnetic susceptibility below {\tn}\,{\cite{kk4}}. 
The magnetic structure has quite similar characteristics to phase IV in {\hbc} 
which is the AFM phase adjacent to the AFQ ordered phase\,{\cite{kk6}}. 
Moreover, the magnetization processes that show multi-step field-induced transitions are very similar to those in {\dbc} and {\hbc}. 
These unusual properties suggest strong AFQ interactions also in {\tbbc}. 
A purpose of this study is to clarify the $H$-$T$ phase diagram of {\tbbc} to shed light on this strong AFQ interaction in the {\rbc} system. 
In this paper, we will report the existence of field-induced AFQ phase and its remarkable stability against magnetic fields. 
{\tbbc} is the first compound which exhibits the field-induced AFQ ordering.
Furthermore, we mention that the phase diagrams in the {\rbc} system are systematically understood 
in terms of the competing AFQ and AFM interactions.

\begin{figure}
\includegraphics[width=7cm]{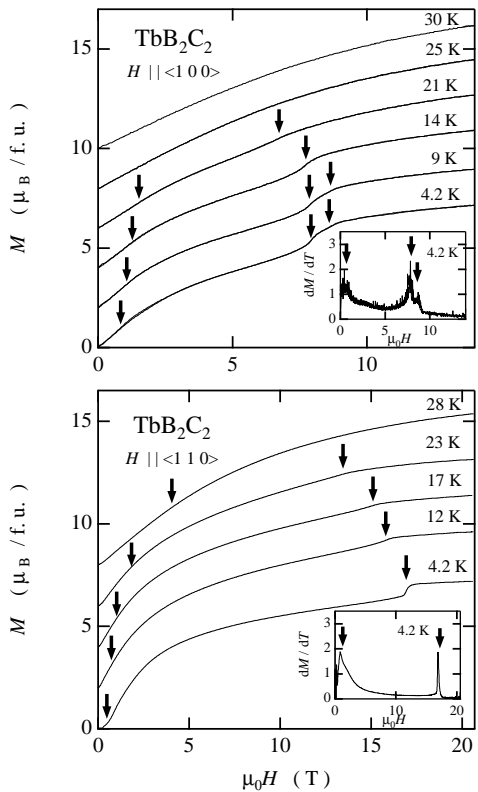}
\caption{\label{fig:MH} Magnetization curves at various temperatures of {\tbbc} under magnetic fields along the 
${\langle}1\,0\,0{\rangle}$ and ${\langle}1\,1\,0{\rangle}$ directions. 
The inset shows the d$M$/d$H$ curve at 4.2\,K for each direction. 
Curves at higher temperatures are raised by certain amounts for clarity.
Each arrow indicates a critical field defined as a maximum in differential magnetization curves.}
\end{figure}%
 
\begin{figure}
\includegraphics[width=7cm]{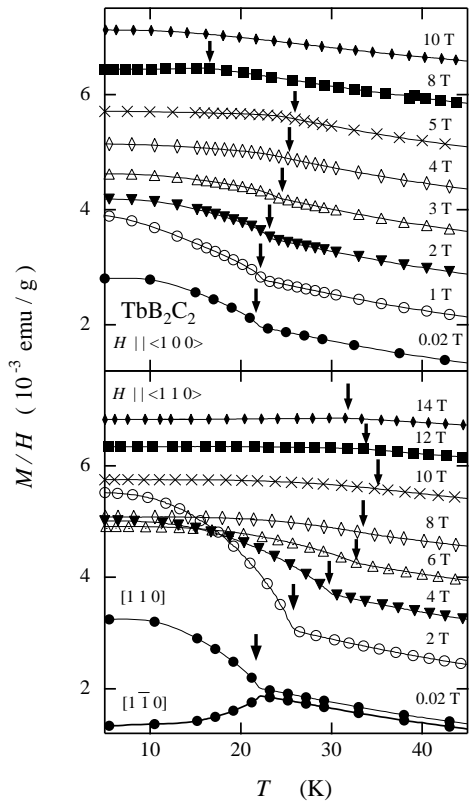}
\caption{\label{fig:MT} Temperature dependence of the magnetization of {\tbbc} under various fields along the 
${\langle}1\,0\,0{\rangle}$ and ${\langle}1\,1\,0{\rangle}$ directions.
Curves at higher fields are raised by certain amounts in order to avoid overlapping of the data.}
\end{figure}%

For sample preparation, we used stoichiometric amounts of the constituents,
Tb of 99.9\%, B of 99.8\% and C of 99.999\% in purity.  
The compound was synthesized through the conventional argon arc melting.  
Single crystalline sample of {\tbbc} was grown by the Czochralski pulling method using a tri-arc furnace. 
The magnetization was measured by using a SQUID magnetometer (Quantum Design) and 
vibrating sample magnetometers with a superconducting magnets of up to 14\,T (Oxford Instruments) and 
a water-cooled Bitter-type steady field magnet up to 15 T installed at High Field Laboratory for 
Superconducting Materials (HFLSM) of Institute for Materials Research (IMR), Tohoku University.
The magnetization processes in higher fields up to 30\,T were measured using a pulse magnet 
and a hybrid-type steady field magnet installed at HFLSM of IMR, Tohoku University. 

\begin{figure*}
\includegraphics[width=12cm]{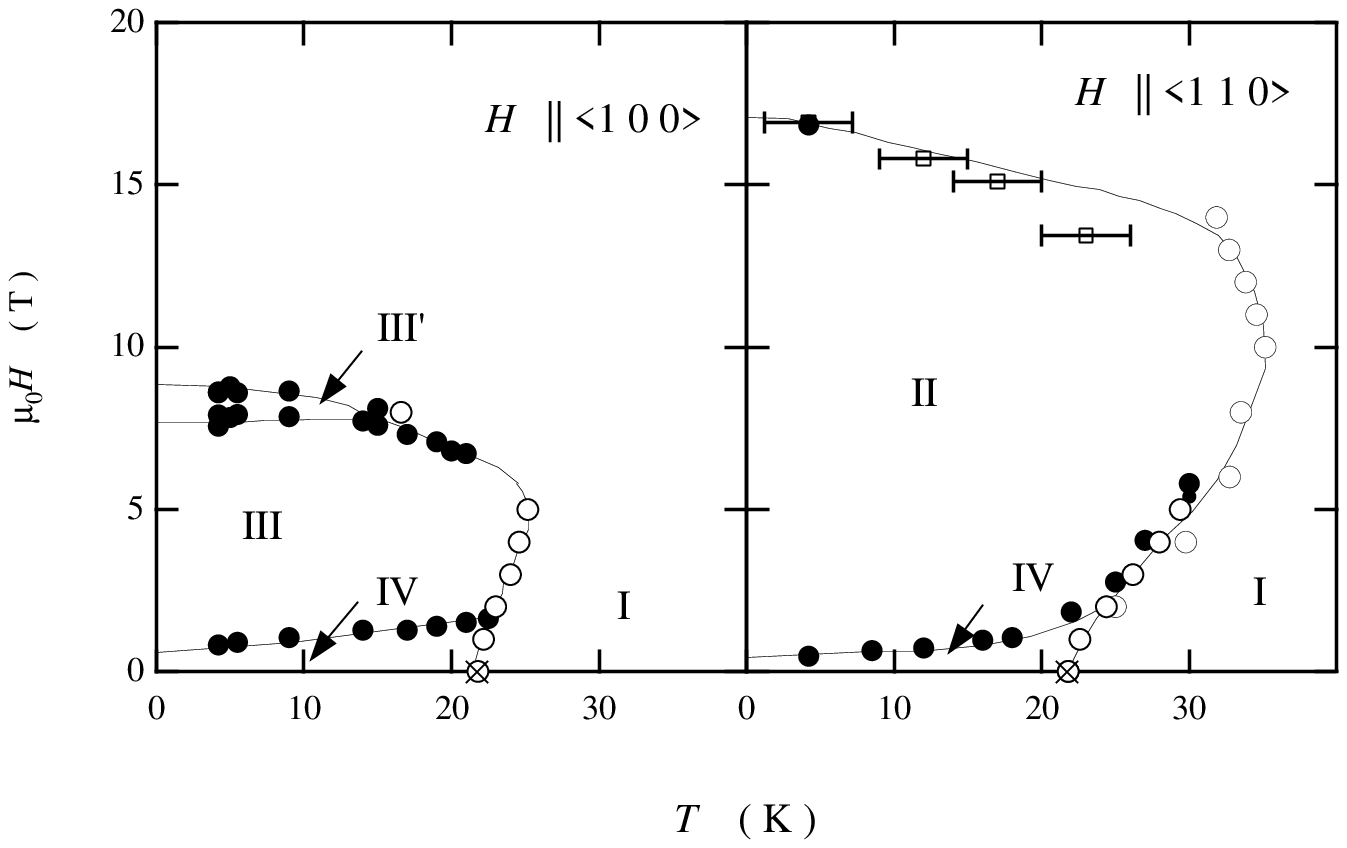}
\caption{\label{fig:Phase}Magnetic $H$-$T$ phase diagrams of {\tbbc} for the {\ea} and {\eab} directions. 
Closed and open circles indicate phase boundaries determined from the 
magnetization processes and the temperature dependence of magnetization. 
Open squares with large errors were obtained from the result of $M$-$H$ curve using pulse magnet.
The N{\'{e}}el temperature determined by specific heat {\cite{kk3} is shown by the crosses.}}
\end{figure*}%

Figure\,{\ref{fig:MH}} shows magnetization processes of {\tbbc} measured at various temperatures.
The insets show the differential d$M$/d$H$ curves at 4.2\,K as functions of magnetic field.     
For $H{\parallel}{\langle}1\,0\,0{\rangle}$, 
two successive transitions were observed at $H$=7.8 and 8.6 T at 4.2\,K.
In addition, as clear in the d$M$/d$H$ curve, a broad anomaly was found around 0.8\,T as well.
With increasing temperature, the transition at the lowest field becomes broader and  
shifts to 1.65\,T at 22.5\,K.
This anomaly becomes unclear above 25 K.
In contrast, the transitions at 7.8 and 8.6\,T shift to lower fields with increasing temperature.
These two transition fields approach and merge into a single anomaly at 7.73\,T at 14\,K.
This transition field decreases rapidly and  no anomaly was observed above 25\,K.

In case for the {\eab} field direction, a broad transition around  0.5\,T 
together with a sharp anomaly at 16.9\,T was observed at 4.2\,K. 
As temperature increases, the lower transition field increases, 
whereas the higher transition field decreases and disappears at 28\,K.
The magnetization process for {\Hdir}[{\rab}] is identical to that for [{\rba}] above 0.4\,T,
although the [{\rab}] axis is inequivalent to [{\rba}] in the ground state due to AFM domains\,{\cite{kk3,kk4}}.
The existence of low field transition is more clearly confirmed by our recent neutron diffraction experiments\,{\cite{kk5}}.

Figure{~\ref{fig:MT}} shows temperature dependence of the magnetization $M/H$ 
at various magnetic fields along the {\ea} and {\eab} directions. 
It should be pointed out that the magnetization increases below {\tn} though {\tbbc} is an antiferromagnet.
In general, the magnetic susceptibility of an antiferromagnet decreases or constant below {\tn}.
This increase was clearly observed up to 4\,T and 11\,T for {\Hdir}{\ea} and {\eab}, respectively.
At higher fields of 14\,T for {\Hdir}{\eab}, the magnetization turns to decrease below {\tn}.
The unusual increase of magnetization is more prominent in {\Hdir}{\eab} than {\ea}.
In case for {\Hdir}[{\rba}], a cusp-like anomaly was observed under 0.02\,T.
This anisotropy is also due to AFM domains{\cite{kk3,kk4}}.
However, the increase  of magnetization below {\tn} was also observed for {\Hdir}[{\rba}] above 2\,T. 

The transition temperature is defined by the temperature of 
the maximum or minimum of ${\rm d}(M/H)/{\rm d}T$ curves as shown by the arrows in Fig.{~\ref{fig:MT}}.
For {\Hdir}{\ea}, we can clearly observe that the transition temperature increases monotonously from 21.7 K
to 25.2\,K at 5\,T with increasing magnetic fields.
In general, the AFM transition temperature should be decreased with the application of magnetic fields.
At higher fields, the transition temperature suddenly decreases.
The fact that no distinct anomaly was observed in the $M/H$ curve at $H=10\,{\rm T}$ indicates that the phase boundary closes between 8 and 10\,T. 

The unusual behavior that the {\tn} increases with the application of fields is 
further remarkable for {\Hdir}{\eab}.
The transition at {\tn}$=21.7$\,K shifts drastically to the higher temperature as field increases, 
and takes the maximum of 35.2\,K at 10\,T.
The transition temperature shows a gradual decrease above 12 T and becomes 31.8\,K at 14\,T.
The close of the phase boundary was not confirmed in this measurement due to the limitation of external field. 

The $H$-$T$ magnetic phase diagrams of {\tbbc} for {\Hdir}{\ea} and {\eab} are shown in Fig.\,{\ref{fig:Phase}}.
The phase I and IV represent the paramagnetic and AFM states, respectively.
Our neutron diffraction study revealed that 
the magnetic structure in phase IV can be described with the propagation vectors of 
a dominant {\kpb} and an additional {\kpd} with a longitudinal sinusoidal modulation  
{\boldmath $k$}$_L=(1\pm{\delta}\,\pm{\delta},0)$ where ${\delta}$=0.13\,{\cite{kk3}}. 
This long periodic modulation {\boldmath $k$}$_L$ is consistent with phase IV of {\hbc} 
which is adjacent to the AFQ phase\,{\cite{kk6}}.  
The most remarkable feature in the magnetic phase diagram of {\tbbc} is 
the existence of the field-induced phases (II and III) which are stable in a wide field range from ${\sim}$0.8\,T 
to 9\,T ({\Hdir}{\ea}) and 17\,T ({\Hdir}{\eab}).
We can clearly recognize that the magnetic field cause the unusual enhancement of the AFM transition temperature
as much as 13.5\,K for the {\eab} direction in Fig.\,{\ref{fig:Phase}}.

In order to identify the phase II and III, the magnetization processes of {\tbbc} were compared with those of {\dbc} and {\hbc}. 
Figure\,{\ref{fig:MH_HM1a} shows the magnetization processes of {\tbbc} at 4.2\,K together with 
those of {\dbc}{\cite{dy1}} and {\hbc}{\cite{Ho_1,hd}} at 1.5\,K.
The applied magnetic field was normalized with the critical field $H_{\rm c}$ for {\Hdir}{\eab} for comparison.
As clearly seen, the magnetization curves of {\tbbc} are very similar to those of {\dbc} and {\hbc}.
In the latter compounds, there is a very wide field region where the AFQ phase II is stable,
when the field is parallel to the {\eab} direction. (See the lower panel in Fig.\,{\ref{fig:MH_HM1a}.)
Therefore, we tentatively assign that the field induced phase in {\tbbc} for {\Hdir}{\eab} would be the AFQ phase.
In lower fields, {\dbc} and {\hbc} show two-step transitions leading to the phase III and III$^{\prime}$, 
in which the AFQ and AFM order coexists.
The difference between phase III and III$^{\prime}$ is most probably the periodicity along the [{\rc}] direction in their magnetic structure\,{\cite{dy3}}.
On the other hand, the two-step transition was not clearly observed in {\tbbc}. 
Thus it is indistinct from the present magnetization measurement
whether the phase III and III$^{\prime}$ exist in {\tbbc} for {\Hdir}{\eab}. 

When the field is applied along the {\ea} direction, phase III, the coexistent phase of AFQ and AFM order, 
shows remarkable stability against magnetic fields.
The phase III is transformed into the paramagnetic state through the intermediate phase III$^{\prime}$ in {\dbc},
while the phase III of {\hbc} directly undergoes magnetic transition to phase I around $H/H_{\rm c}{\sim}$0.5.
{\tbbc} also exhibits the successive transitions around $H/H_{\rm c}{\sim}$0.5.
Therefore, we suggest that the field-induced phase in {\tbbc} from $H/H_{\rm c}{\sim}$0.047 to 0.46 would be phase III.
Furthermore, the successive transition around $H/H_{\rm c}{\sim}$0.5 would be indicative of 
the existence of the very narrow intermediate phase III$^{\prime}$.
Neutron diffraction experiments under magnetic fields are highly interesting to confirm the field-induced AFQ phase in {\tbbc}. 
The existence of AFQ ordering in phase II and III stabilized with magnetic field could be interpreted that 
the strong AFQ interaction does exist in {\tbbc} as well. 

The $H$-$T$ phase diagrams in the {\rbc} system can be understood in terms of competition of the AFQ and AFM interactions.
The most distinct character in {\dbc} is the existence of the pure AFQ phase II at $H$=0 without any magnetic contribution.
It could be understood that the AFQ interaction is strong enough that the AFQ ordering survives as an intermediate phase to the paramagnetic phase. 
In case of {\hbc}, however, the pure AFQ phase can only be stabilized in finite magnetic fields.
For $H$=0, the AFQ ordering is accompanied with the AFM ordering.
In {\tbbc}, the AFM interaction does not allow AFQ ordering under zero magnetic field, and the AFQ ordering is only realized under magnetic fields. 
It should be noted, however, that the AFM phase (phase IV) lies adjacent to the AFQ phase 
and is strongly affected by the AFQ interaction.

\begin{figure}
\includegraphics[width=7.0cm]{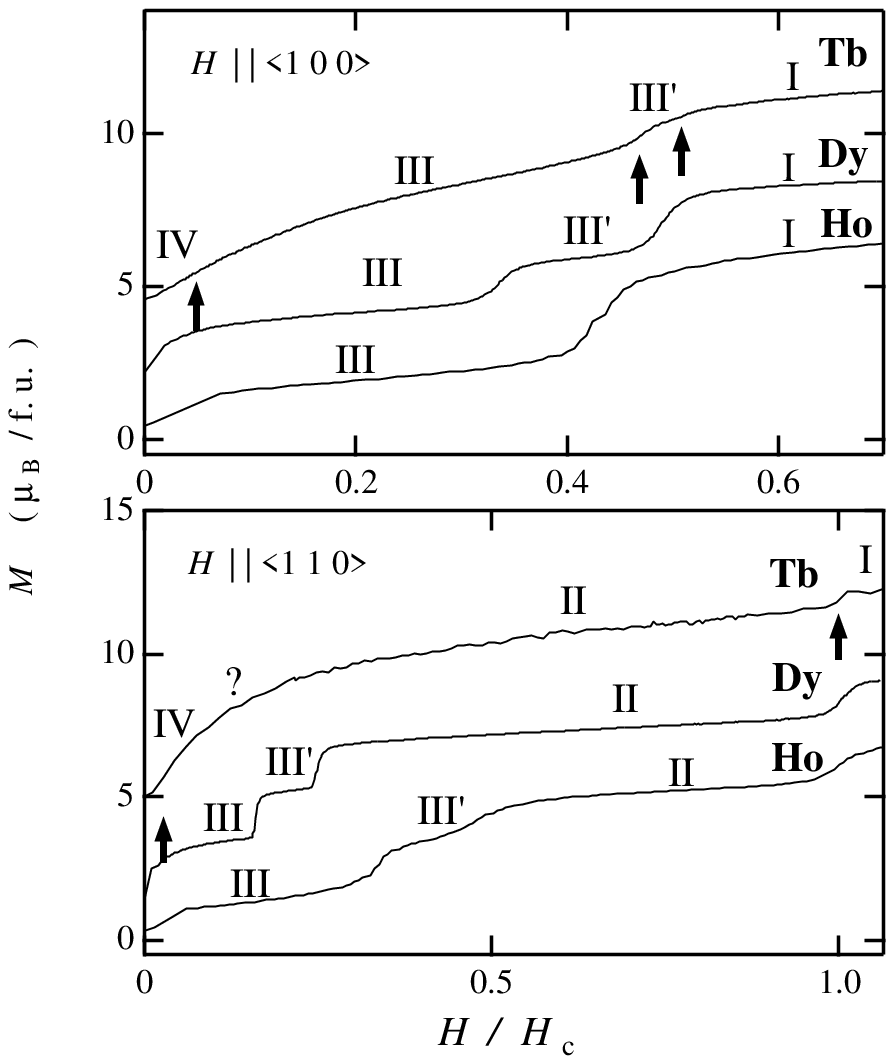}
\caption{\label{fig:MH_HM1a} Magnetization processes of {\tbbc} (4.2\,K) and {\dbc}\,{\cite{dy1}} and {\hbc}\,{\cite{Ho_1}}\,(1.5\,K)
for {\Hdir}{\ea} and {\eab}. The applied magnetic fields are normalized by the highest transition field in the {\eab} direction of each compound. }
\end{figure}%

The unusual enhancement of {\tn} with magnetic field is a distinctive character in {\tbbc}.
The increase of {\tn} in {\tbbc} reaches 13.5 K which corresponding to ${\sim}$1\,K/T.
This enhancement is quite large in comparison with {\dbc} and {\hbc}. 
In these compounds, the AFQ transition temperature increases relatively small about 1\,K.  
The increase of {\tq} was also reported for other AFQ materials {\cb}\,{\cite{ceb6_0,ceb6}} and {\pb}\,{\cite{prpb_1}}. 
In {\pb} with non-magnetic ${\Gamma}_3$ ground state,
the interaction between field-induced staggered moments stabilizes the AFQ order.
However, the increase of the transition temperature reaches only 0.3\,K at 6\,T. 
With respect to {\cb}, the AFQ transition temperature {\tq}=3.3\,K is raised to 9.5\,K by the external field of 30\,T\,{\cite{ceb6_5}}.
Recent theoretical works succeeded to explain this anomalous increase of {\tq} in {\cb}
by taking octupolar interaction into account\,{\cite{ceb6_2,ceb6_1,ceb6_4,ceb6_3}}. 
An antiferro-type interaction between field-induced octupoles stabilizes the AFQ order against magnetic fields. 
It is, therefore, supposed that the octupolar moments in {\tbbc} 
 have an important role in the magnetic behavior than that in {\dbc} and {\hbc}.

In conclusion, the antiferromagnet {\tbbc} is under the strong influence of AFQ interactions. 
We suggest that {\tbbc} is the first compound which shows the field-induced AFQ ordering.
The comparison of the magnetic phase diagrams of {\tbbc} with those of {\dbc} and {\hbc} 
indicates the essential role of competing AFQ and AFM interactions in {\rbc} system. 

The authors would like to thank N. Metoki for stimulating discussions.
The staff of HFLSM of IMR is gratefully acknowledged for operation of the high-field magnets.
This work was supported partially by a Grant-in-Aid for Scientific Research
(No. 12304017) from the Japan Society for the Promotion of Science.

\end{document}